\documentstyle[12pt]{article}
\def\case#1#2{{\textstyle{#1\over #2}}}

\title{
\hfill{\normalsize ULB/229/CQ/99/2}\\
\vspace{1cm}
Comment on `Application of nonlinear deformation algebra to a physical system
with P\"oschl-Teller potential'}
\author{C Quesne\thanks{Directeur de recherches FNRS} \thanks{E-mail address:
cquesne@ulb.ac.be}\\
{\small Physique Nucl\'eaire Th\'eorique et Physique Math\'ematique,}\\
{\small Universit\'e Libre de Bruxelles, Campus de la Plaine CP229,}\\
{\small Boulevard~du Triomphe, B-1050 Brussels, Belgium}}
\date{ }
\begin{document}
\baselineskip=20pt plus 1pt minus 1pt
\maketitle

\begin{abstract}
We comment on a recent paper by Chen, Liu, and Ge ({\em J.\ Phys.\ A:
Math.\ Gen.}
{\bf 31} (1998) 6473), wherein a nonlinear deformation of su(1,1) involving two
deforming functions is realized in the exactly solvable quantum-mechanical
problem with P\" oschl-Teller potential, and is used to derive the well-known
su(1,1) spectrum-generating algebra of this problem. We show that one of the
defining relations of the nonlinear algebra, presented by the authors, is
only valid
in the limiting case of an infinite square well, and we determine the correct
relation in the general case. We also use it to establish the correct link with
su(1,1), as well as to provide an algebraic derivation of the eigenfunction
normalization constant.
\end{abstract}

\bigskip\bigskip
Short title: Application of nonlinear deformation algebra

\newpage
%
%
In an interesting paper (henceforth referred to as I and whose equations will be
quoted by their number preceded by I), Chen, Liu, and Ge~\cite{chen} recently
pointed out that the nonlinear deformations of the su(2) and su(1,1) Lie
algebras
with two deforming functions $f(J_0)$ and $g(J_0)$, introduced by Delbecq and
Quesne~\cite{delbecq}, can find some useful applications in quantum mechanics.
They indeed claim to have proved that one of such algebras can be realized in a
physical system with P\" oschl-Teller potential, which is one of the exactly
solvable one-dimensional quantum-mechanical potentials.\par
%
%
By starting from the `natural' quantum operators $X$, $P$ of Nieto and
Simmons~\cite{nieto79}, they constructed mutually adjoint lowering and raising
operators $b$, $b^+$, which together with the Hamiltonian $H$ generate a
nonlinear
algebra with two deforming functions $f(H)$ and $g(H)$. They also obtained the
eigenvalues and (unnormalized) eigenfunctions of $H$ by using this algebra
instead
of solving the Schr\" odinger equation, and pointed out a relation with the
well-known su(1,1) symmetry of the P\" oschl-Teller potential (see~\cite{levai}
ans references quoted therein).\par
%
%
In the present comment, we want to show that one of the defining relations
of the
nonlinear algebra, as given in~I, is not entirely correct, and should
actually contain
an additional term, which only disappears in the $\nu \to 1$ limit,
corresponding to
an infinite square well. In support of the amended relation, we will prove
that it
allows us to algebraically derive the known eigenfunction normalization
constant~\cite{nieto78}. Finally, we will establish the correct relation between
the nonlinear algebra and su(1,1).\par
%
%
Let $H$, $b$, $b^+$ be defined as in~I by
\begin{equation}
  H = \frac{p^2}{2m} + V(x) \qquad V(x) = \frac{V_0}{\cos^2(kx)} \qquad
  V_0 = \epsilon \nu (\nu-1) \qquad \epsilon = \frac{\hbar^2 k^2}{2m}
  \label{eq:H}
\end{equation}
\begin{equation}
  b = \frac{1}{2\epsilon} \left[X \left(\epsilon + 2 \sqrt{\epsilon H}\right)
  + \frac{{\rm i}\hbar}{m} P\right]  \label{eq:b}
\end{equation}
\begin{equation}
  b^+ = (b)^{\dagger} = \frac{1}{2\epsilon} \left[\left(\epsilon + 2
\sqrt{\epsilon
  H}\right) X - \frac{{\rm i}\hbar}{m} P\right] = - \frac{1}{2\epsilon}
  \left[X \left(\epsilon - 2 \sqrt{\epsilon H}\right) + \frac{{\rm i}\hbar}{m}
  P\right] \frac{\epsilon + \sqrt{\epsilon H}}{\sqrt{\epsilon H}}
\label{eq:bplus}
\end{equation}
where
\begin{equation}
  X = \sin(kx) \qquad P = \case{1}{2} k \{\cos(kx), p\} = k \left[\cos(kx) p
  + \case{1}{2} {\rm i} \hbar k \sin(kx)\right]  \label{eq:XP}
\end{equation}
satisfy the commutation relations
\begin{eqnarray}
  [X, P] & = & {\rm i} \hbar k^2 \left(1 - X^2\right) \qquad [H, X] = -
\frac{{\rm i}
  \hbar}{m} P \nonumber \\[0cm]
  [H, P] & = & {\rm i} \hbar k^2 \left(2XH - \frac{1}{2} \epsilon X -
  \frac{{\rm i} \hbar}{m} P\right).  \label{eq:XP-com}
\end{eqnarray}
Note that in equations~(\ref{eq:b}) and~(\ref{eq:bplus}), we have set $\gamma =
(2\epsilon)^{-1}$ in accordance with equation~(I45), and that $\nu$ can be
expressed in terms of $V_0$ as
\begin{equation}
  \nu = \frac{1}{2} \left(1 + \sqrt{1 + \frac{4V_0}{\epsilon}}\right).
\label{eq:nu}
\end{equation}
Equation~(\ref{eq:nu}) differs from equation~(I35), wherein a minus sign is used
before the square root. The plus sign is actually imposed by the condition
that the
wave functions vanish at the boundaries of the interval $\left(- \frac{\pi}{2k},
\frac{\pi}{2k}\right)$.\par
%
%
{}For the commutators of $H$, $b$, and $b^+$, we obtain the results
\begin{eqnarray}
  [H, b] & = & - b g(H) \qquad \left[H, b^+\right] = g(H) b^+ \qquad g(H) =
- \epsilon
           + 2 \sqrt{\epsilon H}  \label{eq:H-b-bplus} \\
  \left[b, b^+\right] & = & - f(H) = 1 + 2 \sqrt{H/\epsilon} + \frac{\nu(\nu-1)}
           {\sqrt{H/\epsilon}\left(\sqrt{H/\epsilon}-1\right)}
\label{eq:b-bplus}
\end{eqnarray}
which only partly agree with equations~(I17) and~(I18), as the last term on the
right-hand side of equation~(\ref{eq:b-bplus}) is missing there. Since the
calculation of the commutator of $b$ with $b^+$ is not quite straightforward, we
shall now provide some steps of the proof of equation~(\ref{eq:b-bplus}).\par
%
%
By using equations~(\ref{eq:b}) and~(\ref{eq:bplus}), as well as the method
devised
in~I for commuting a function of $H$ with $X$ or $P$, we get
\begin{eqnarray}
  b b^+ & = & - \frac{1}{4\epsilon^2} \Biggl\{\left[X^2 \left(\epsilon -
         2\sqrt{\epsilon H}\right) + \frac{{\rm i}\hbar}{m} XP\right]
         \left(3\epsilon + 2\sqrt{\epsilon H}\right) + \frac{{\rm i}\hbar}{m} PX
         \left(\epsilon - 2\sqrt{\epsilon H}\right) \nonumber \\
  & & \mbox{} - \frac{\hbar^2}{m^2} P^2\Biggr\} \frac{\epsilon + \sqrt{\epsilon
         H}}{\sqrt{\epsilon H}} \\
  b^+ b & = & \frac{1}{4\epsilon^2} \Biggl\{\left[X^2 \left(\epsilon +
         2\sqrt{\epsilon H}\right) + \frac{{\rm i}\hbar}{m} XP\right]
         \left(3\epsilon - 2\sqrt{\epsilon H}\right) + \frac{{\rm i}\hbar}{m} PX
         \left(\epsilon + 2\sqrt{\epsilon H}\right) \nonumber \\
  & & \mbox{} - \frac{\hbar^2}{m^2} P^2\Biggr\} \frac{\sqrt{\epsilon
H}}{\epsilon -
         \sqrt{\epsilon H}}.
\end{eqnarray}
From these results and equation~(\ref{eq:XP-com}), we obtain the rather
complicated expression
\begin{eqnarray}
  \left[b, b^+\right] & = & \frac{1}{4\epsilon} \Biggl\{-2 \left[\epsilon^3 -
        2 \epsilon^2 (\epsilon H)^{1/2} + 4 (\epsilon H)^{3/2}\right] - X^2
\epsilon
        \left(\epsilon^2 - 4 \epsilon H\right) - 4 \frac{{\rm i}\hbar}{m}
\epsilon^2 XP
        \nonumber \\
  & & \mbox{} + \frac{\hbar^2}{m^2} \epsilon P^2\Biggr\} \frac{1}
{\sqrt{\epsilon
        H} \left(\epsilon - \sqrt{\epsilon H}\right)}.
\label{eq:b-bplus-interm}
\end{eqnarray}
The latter can however be simplified by noting that
\begin{eqnarray}
  \nu (\nu-1) & = & \frac{1}{\epsilon} \left(1 - X^2\right) \left(H -
\frac{p^2}{2m}
         \right) \nonumber \\
  & = & \frac{1}{4\epsilon^2} \left[2 \left(\epsilon^2 + 2 \epsilon
H\right) + X^2
         \left(\epsilon^2 - 4 \epsilon H\right) + 4 \frac{{\rm i}\hbar}{m}
\epsilon XP
         - \frac{\hbar^2}{m^2} P^2\right]  \label{eq:Vzero}
\end{eqnarray}
where the first equality directly results from equation~(\ref{eq:H}). To
obtain the
second equality, use has been made of equation~(\ref{eq:XP}), leading to
\begin{equation}
  [\cos(kx) p]^2 = \left(- \frac{1}{2} {\rm i} \hbar k X + \frac{P}{k}\right)^2
\end{equation}
or
\begin{equation}
  \left(1 - X^2\right) p^2 + \frac{1}{2} \hbar^2 k^2 X^2 + {\rm i} \hbar X P =
  - \frac{1}{2} \hbar^2 k^2 + \frac{1}{4} \hbar^2 k^2 X^2 - {\rm i} \hbar X P
  + \frac{1}{k^2} P^2.
\end{equation}
Inserting equation~(\ref{eq:Vzero}) into equation~(\ref{eq:b-bplus-interm})
completes the proof of equation~(\ref{eq:b-bplus}).\par
%
%
We conclude that $J_0 = H$, $J_+ = b^+$, and $J_- = b$ do indeed define a
nonlinear algebra with two deforming functions $f(H)$ and $g(H)$, as introduced
in~\cite{delbecq}, but that $f(H)$ does contain an extra term not obtained in~I.
Since such a term depends upon the potential strength, we get different algebras
for different Hamiltonians. It should be noted that the presence of
$\sqrt{H/\epsilon}$ in the denominator does not lead to any problem when
acting on
the Hamiltonian eigenstates.\par
%
%
The Casimir operator of the nonlinear algebra is given by~\cite{delbecq}
\begin{equation}
  C = b b^+ + h(H) = b^+ b + h(H) - f(H)  \label{eq:casimir}
\end{equation}
where $h(H)$ satisfies the relation $h(H) - h(H - g(H)) = f(H)$, and is given by
\begin{equation}
  h(H) = - \left(1 + \sqrt{H/\epsilon}\right)^2 + \frac{\nu(\nu-1)}
  {\sqrt{H/\epsilon}}.  \label{eq:h}
\end{equation}
This result, too, differs from equations~(I2) and~(I19).\par
%
%
The nonlinear algebra can be used to determine the spectrum of $H$ and to
construct all its eigenstates, and is therefore a spectrum-generating
algebra. As
proved in~I, the eigenvalues and the ground state wave function, obtained
from the
equation
\begin{equation}
  b |\psi_0\rangle = 0  \label{eq:groundstate}
\end{equation}
are given by
\begin{equation}
  E_n = \epsilon (n + \nu)^2 \qquad n = 0, 1, 2, \ldots  \label{eq:eigenvalues}
\end{equation}
and
\begin{equation}
  \psi_0(x) = {\cal N}_0 \cos^{\nu}(kx)
\end{equation}
respectively. The normalization constant ${\cal N}_0$, calculated by direct
integration, is
\begin{equation}
  {\cal N}_0 = \left(\frac{k \Gamma(\nu+1)}{\sqrt{\pi}\, \Gamma(\nu+1/2)}\right)
  ^{1/2}.  \label{eq:norm-zero}
\end{equation}
\par
%
%
The excited states $|\psi_n\rangle$, $n=1$, 2,~\ldots, can be obtained by
repeatedly using the relation
\begin{equation}
  b^+ |\psi_n\rangle = \alpha_{n+1} |\psi_{n+1}\rangle  \label{eq:bplus-action}
\end{equation}
where $\alpha_{n+1}$ is some yet unknown constant, which we may assume real
and nonnegative. By using the explicit form of $b^+$ and the Hamiltonian
eigenvalues, given in equations~(\ref{eq:bplus}) and~(\ref{eq:eigenvalues}),
respectively, we get the recursion relation
\begin{equation}
  \frac{n+\nu+1}{n+\nu} \left[- \frac{1}{k} \cos(kx) \frac{d}{dx} + (n+\nu)
\sin(kx)
  \right] \psi_n(x) = \alpha_{n+1} \psi_{n+1}(x).  \label{eq:psi-recursion}
\end{equation}
If we set
\begin{equation}
  \psi_n(x) = \left(1 - X^2\right)^{\nu/2} \phi_n(X)  \label{eq:psi}
\end{equation}
where $X$ is defined in equation~(\ref{eq:XP}), then
equation~(\ref{eq:psi-recursion}) becomes
\begin{equation}
  \left[\left(X^2 - 1\right) \frac{d}{dX} + (n+2\nu) X\right] \phi_n(X) =
\frac{n+\nu}
  {n+\nu+1} \alpha_{n+1} \phi_{n+1}(X).
\end{equation}
Comparison with the differential and recursion relations of Gegenbauer
polynomials~\cite{abramowitz} shows that
\begin{equation}
  \phi_n(X) = {\cal N}_n C^{(\nu)}_n(X)  \label{eq:phi}
\end{equation}
where the normalization constant ${\cal N}_n$ satisfies the recursion relation
\begin{equation}
  \frac{{\cal N}_n}{{\cal N}_{n-1}} = \frac{n (n+\nu)}{(n+\nu-1) \alpha_n}.
  \label{eq:norm-recursion}
\end{equation}
Equations~(\ref{eq:psi}) and~(\ref{eq:phi}), which were already obtained
before by
Eleonsky and Korolev~\cite{eleonsky}, are equivalent to equations~(I39)
and~(I40),
but provide a simpler expression for the wave functions.\par
%
%
To get the wave-function normalization constant ${\cal N}_n$, it is clear from
equation~(\ref{eq:norm-recursion}) that we need an explicit expression for
$\alpha_n$. By considering the diagonal matrix element (with respect to
$|\psi_n\rangle$) of both sides of equation~(\ref{eq:b-bplus}), and using
equation~(\ref{eq:eigenvalues}), as well as the relation
\begin{equation}
  b |\psi_n\rangle = \alpha_n |\psi_{n-1}\rangle  \label{eq:b-action}
\end{equation}
we obtain the following recursion relation for $|\alpha_n|^2$,
\begin{equation}
   |\alpha_n|^2 - |\alpha_{n-1}|^2 = 2n + 2\nu - 1 + \frac{\nu
(\nu-1)}{(n+\nu-1)
  (n+\nu-2)}.
\end{equation}
Its solution is given by
\begin{equation}
  |\alpha_n|^2 = (n+\nu)^2 - \frac{\nu (\nu-1)}{n+\nu-1} + \beta
\end{equation}
where $\beta$ is some constant. Since equation~(\ref{eq:groundstate})
imposes the
condition $\alpha_0 = 0$, we get $\beta = - \nu (\nu-1)$. Hence
\begin{equation}
  \alpha_n = \left(\frac{n (n+\nu) (n+2\nu-1)}{n+\nu-1}\right)^{1/2}
  \label{eq:alpha}
\end{equation}
in accordance with equations (3.30) and (3.31) of~\cite{nieto79}.\par
%
%
Inserting equation~(\ref{eq:alpha}) into equation~(\ref{eq:norm-recursion}) now
leads to the result
\begin{equation}
  {\cal N}_n = {\cal N}_0 \left(\frac{n!\, (n+\nu) \Gamma(2\nu)}{\nu\,
  \Gamma(n+2\nu)}\right)^{1/2}.  \label{eq:norm}
\end{equation}
By combining equations~(\ref{eq:norm-zero}), (\ref{eq:psi}), (\ref{eq:phi}),
and~(\ref{eq:norm}), we obtain the following form for the normalized wave
functions
\begin{equation}
  \psi_n(x) = \left(\frac{k (n!) (n+\nu) \Gamma(\nu) \Gamma(2\nu)}{\sqrt{\pi}\,
  \Gamma(\nu+1/2) \Gamma(n+2\nu)}\right)^{1/2} \cos^{\nu}(kx) C^{(\nu)}_n
  (\sin(kx)).
\end{equation}
It only remains to take the known relation between Gegenbauer polynomials and
associated Legendre functions~\cite{abramowitz} into account to get the
equivalent form
\begin{equation}
  \psi_n(x) = \left(\frac{k (n+\nu) \Gamma(n+2\nu)}{n!}\right)^{1/2}
\cos^{1/2}(kx)
  P^{1/2-\nu}_{n+\nu-1/2}(\sin(kx))
\end{equation}
given by Nieto~\cite{nieto78}.\par
%
Equation~(\ref{eq:alpha}), together with equations~(\ref{eq:eigenvalues}),
(\ref{eq:bplus-action}), and~(\ref{eq:b-action}), also allows us to
determine the
eigenvalue of the Casimir operator~(\ref{eq:casimir}) corresponding to
$|\psi_n\rangle$,
\begin{equation}
  C |\psi_n\rangle = - \nu (\nu-1) |\psi_n\rangle.
\label{eq:casimir-eigenvalue}
\end{equation}
Hence, all the Hamiltonian eigenstates $|\psi_n\rangle$ belong to a single
unitary
irreducible representation of the nonlinear algebra, which may be characterized
by~$\nu$.\par
%
%
At this stage, we may transform the nonlinear algebra in two different
ways: either
by trying to free ourselves from the need for considering different algebras for
different Hamiltonians, or by restricting ourselves to the irreducible
representation wherein equation~(\ref{eq:casimir-eigenvalue}) is satisfied.\par
%
%
In the former case, we may use equations~(\ref{eq:casimir}) and~(\ref{eq:h}) to
express $\nu(\nu-1)$ in terms of $C$, $bb^+$, and $\sqrt{H/\epsilon}$. Inserting
such an expression into equation~(\ref{eq:b-bplus}) then leads to an extended
nonlinear algebra, generated by $H$, $b$, $b^+$, $C$, and characterized by the
defining relations~(\ref{eq:H-b-bplus}), as well as
\begin{equation}
  [C, H] = [C, b] = \left[C, b^+\right] = 0
\end{equation}
\begin{equation}
  \sqrt{H/\epsilon}\, b b^+ - \left(\sqrt{H/\epsilon} - 1\right) b^+ b = C
  + \sqrt{H/\epsilon} \left(1 + 3\sqrt{H/\epsilon}\right).
\end{equation}
Such an algebra may serve as a spectrum-generating algebra for the whole
class of
P\" oschl-Teller Hamiltonians~(\ref{eq:H}).\par
%
%
In the latter case, we may replace $\nu(\nu-1)$ by $-C$ in
equation~(\ref{eq:b-bplus}). Doing the same in equations~(\ref{eq:casimir})
and~(\ref{eq:h}), we may express $C$ in terms of $bb^+$ and
$\sqrt{H/\epsilon}$ as
\begin{equation}
  C = \frac{\sqrt{H/\epsilon}}{\sqrt{H/\epsilon} + 1}\, b b^+ -
\sqrt{H/\epsilon}
  \left(\sqrt{H/\epsilon} + 1\right).  \label{eq:transformed-C}
\end{equation}
Inserting this expression into the transformed equation~(\ref{eq:b-bplus})
leads to
the relation
\begin{equation}
  \frac{\sqrt{H/\epsilon}}{\sqrt{H/\epsilon} + 1}\, b b^+ -
\frac{\sqrt{H/\epsilon} -
  1}{\sqrt{H/\epsilon}}\, b^+ b = 2 \sqrt{H/\epsilon}.
\end{equation}
Hence, the operators
\begin{eqnarray}
  J_0 & = & \sqrt{H/\epsilon}, \qquad J_+ = b^+ \left(\frac{\sqrt{H/\epsilon}}
          {\sqrt{H/\epsilon} + 1}\right)^{1/2} =
\left(\frac{\sqrt{H/\epsilon} - 1}
          {\sqrt{H/\epsilon}}\right)^{1/2} b^+ \nonumber \\
  J_- & = & \left(\frac{\sqrt{H/\epsilon}}{\sqrt{H/\epsilon} + 1}\right)^{1/2} b
          = b \left(\frac{\sqrt{H/\epsilon} - 1}{\sqrt{H/\epsilon}}\right)^{1/2}
          \label{eq:su(1,1)}
\end{eqnarray}
satisfy the defining relations of su(1,1)
\begin{equation}
  [J_0, J_{\pm}] = \pm J_{\pm} \qquad [J_+, J_-] = - 2J_0
\end{equation}
while the operator~(\ref{eq:transformed-C}) reduces to the su(1,1) Casimir
operator, $C = J_- J_+ - J_0 (J_0+1)$. We conclude that under the
substitution of
$-C$ for $\nu(\nu-1)$, the spectrum-generating nonlinear algebra becomes
equivalent to the well-known su(1,1) algebra of the P\" oschl-Teller
potential.\par
%
%
As a final point, let us comment on the limit $V_0 \to 0$ or $\nu \to 1$,
corresponding to an infinite square well of width $L = \pi/k$. In such a
case, the
additional term in equation~(\ref{eq:b-bplus}) vanishes, so that the
results of~I are
applicable. In particular, equation~(I48) provides an acceptable realization of
su(1,1). Since such a realization differs from equation~(\ref{eq:su(1,1)}),
one may
wonder at such a discrepancy. The latter is however easily understood by noting
that with realization~(I48) the nonlinear algebra Casimir operator $C$ actually
differs from that of su(1,1) by an additive constant, whereas with
realization~(\ref{eq:su(1,1)}) both exactly coincide.\par
%
%

\end{document}